\documentclass[aps,prl,reprint,superscriptaddress,showpacs]{revtex4-1}

\usepackage{verbatim}
\usepackage{graphicx}

\newcommand{\esec}[1]{\textit{#1}. }
\newcommand{\NL}{N_{1}}
\newcommand{\NR}{N_{2}}
\newcommand{\minlist}{\vspace{-0.33\baselineskip}}
\newcommand{\fref}[1]{Fig.~\ref{#1}}
\newcommand{\frefs}[1]{Figs.~\ref{#1}}
\newcommand{\eref}[1]{Eq.~(\ref{#1})}
\newcommand{\Tr}{\mathrm{Tr}}
\newcommand{\rmi}{\mathrm{i}}

\begin{document}

\title{Universality in chaotic quantum transport:\\
The concordance between random matrix and semiclassical theories}

\author{Gregory Berkolaiko}
\affiliation{Department of Mathematics, Texas A\&M University, College Station,
TX 77843-3368, USA}
\author{Jack Kuipers}
\affiliation{Institut f\"ur Theoretische Physik, Universit\"at Regensburg,
D-93040 Regensburg, Germany}

\date{\today}

\begin{abstract}
  Electronic transport through chaotic quantum dots exhibits
  universal, system independent, properties, consistent with random
  matrix theory.  The quantum transport can also be rooted, via the
  semiclassical approximation, in sums over the classical scattering
  trajectories. Correlations between such trajectories can be
  organized diagrammatically and have been shown to yield universal
  answers for some observables.  Here, we develop the general
  combinatorial treatment of the semiclassical diagrams, through a
  connection to factorizations of permutations.  We show agreement
  between the semiclassical and random matrix approaches to the
  moments of the transmission eigenvalues.  The result is valid for all
  moments to all orders of the expansion in inverse channel number 
  for all three main symmetry classes (with and without time reversal 
  symmetry and spin-orbit interaction) and extends to nonlinear statistics. 
  This finally explains the applicability of random matrix theory to
  chaotic quantum transport in terms of the underlying dynamics as
  well as providing semiclassical access to the probability density of
  the transmission eigenvalues.
\end{abstract}

\pacs{05.45.Mt, 73.23.-b, 03.65.Nk, 03.65.Sq}

\maketitle

Closed mesoscopic systems with sizes between the atomic and macroscopic possess
statistically different energy spectra depending on whether the corresponding
classical system is regular or chaotic \cite{gutzwiller90,haake10}.  A semiclassical approach
to such systems, valid in the effective limit of $\hbar\to0$, leads to trace
formulae where the density of energy states is approximated by sums over the
classical periodic orbits of the system \cite{gutzwiller71,bb74} which form
stable families for regular systems while being unstable and isolated in chaotic
ones.  A hallmark of the energy statistics is the form factor, a two-point
correlation function approximated by a double sum over periodic orbits.  By
pairing orbits with themselves for chaotic systems or members of their families
for regular ones, the difference between their corresponding energy spectra can
be directly linked to the properties of the underlying dynamics
\cite{ha84,berry85}.  

For (quantum) chaotic systems, there is the further conjecture \cite{bgs84} that
the statistics of the energy spectra are universal (depending just on the
symmetry of the system) and identical to those of the eigenvalues of large
random matrices \cite{mehta04}, originally employed to model the spectra of 
atomic nuclei.  However, the semiclassical pairing of periodic orbits with themselves
\cite{berry85} only led to agreement with the leading order term of
the random matrix theory (RMT) form
factor.  Recently, additional correlated periodic orbits were identified,
treated and shown to provide exact agreement with RMT for short times 
\cite{sr01,mulleretal04,*mulleretal05}.  This involves orbits which come close
to themselves in an `encounter', whose occurrence can be estimated using the
ergodicity of the classical motion, and partner orbits which can be constructed,
due to the local hyperbolicity, to cross the encounter differently.  
For long times the correlations remain unknown, but the form factor can be 
obtained through resummation of short orbits \cite{heusleretal07,km07,mulleretal09}.

Hallmarks of the underlying dynamics persist in open systems, obtained
by attaching scattering leads, as seen for example in an experimental 
study of the electronic transport through quantum dots \cite{changetal94}.
Theoretically, we start with the transmission subblock $t$ of the
scattering matrix connecting asymptotic states in the (two) leads.
The transmission eigenvalues of the matrix
$T=t^{\dagger}t$, and their moments $M_{n}=\Tr [T^{n}]$ relate to the
electronic flow through the system.  For example in the low
temperature limit the first moment is proportional to the conductance
\cite{Landauer57, *Landauer88, Buttiker86}.

\begin{figure}
\includegraphics[width=\columnwidth]{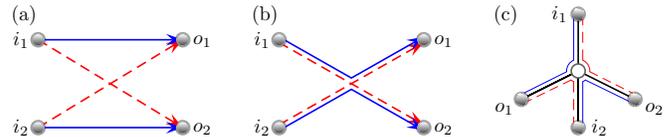}
\caption{\label{semidiags}(Color online) (a) The trajectories for $M_2$ form a
closed cycle if the dashed trajectories, which contribute with negative action,
are traversed backwards.  (b) To contribute in the semiclassical limit, the
trajectories must be nearly identical apart from in small encounter regions
which they can traverse differently.  (c)  By untwisting the encounter, the
diagram  in (b) can be redrawn as the boundary walk of a tree.}
\end{figure}
\begin{figure*}
\includegraphics[width=\textwidth]{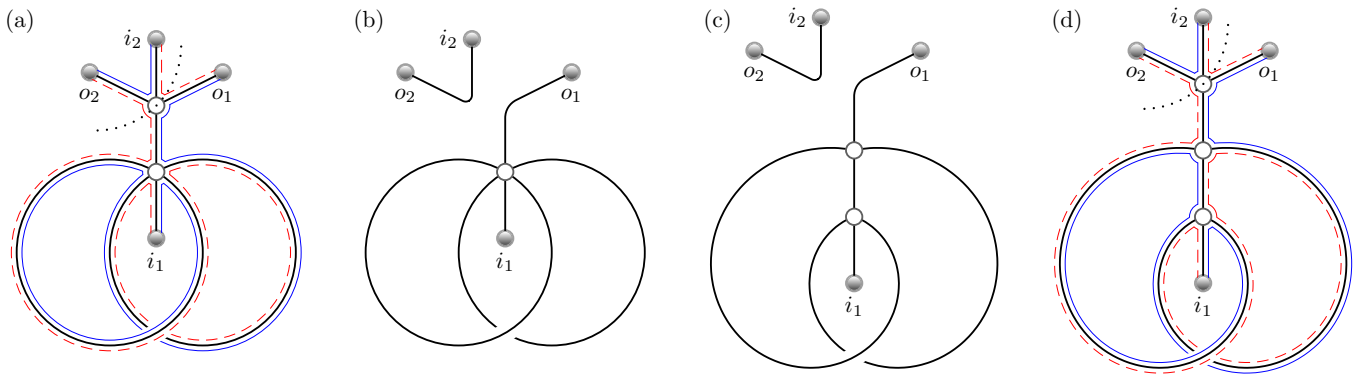}
\caption{\label{graphcancel}(Color online) The graph in (a) and its boundary
walk represent a trajectory quartet which would contribute to $M_2$.  Following
steps 1--5 we first cut the top encounter node (keeping $o_2$ connected to
$i_2$) and arrive at (b).  As $o_1$ is now attached to a node of degree 6 we
insert a new link to obtain (c).  Reconnecting the link from $i_2$ to $o_2$ we
obtain a second trajectory quartet in (d) which exactly cancels the contribution
from (a).  Performing steps 1--5 on (d) reverses the chain to recover (a).}
\end{figure*}

For ballistic chaotic systems, modeling the scattering matrix by a
random matrix from the circular ensembles was proposed and shown to be
consistent with a diagonal semiclassical approach
\cite{BluSmi_prl88,*BluSmi_prl90}.  For the low moments $M_1$ and
$M_2$, all off-diagonal contributions were evaluated in
\cite{rs02,heusleretal06,mulleretal07}, while the calculation of 
general $M_n$, but only for the first several off-diagonal terms, 
were performed in \cite{bhn08, bk11}.  In all
cases, the results agree with RMT.  The purpose of this letter is to
exhibit the mathematical reasons behind this agreement and establish the
general equivalence between semiclassics for open systems and RMT of
the circular ensembles.  The derivation extends to all three main
symmetry classes: unitary for chaotic systems without 
time reversal symmetry (TRS), orthogonal for systems with TRS and 
symplectic for systems with spin-orbit interaction.  
It also remains valid for the nonlinear moments.  

Our approach holds in the universal regime where the dwell 
time, the average time spent inside the system, is much longer than 
the Ehrenfest time $\tau_{\mathrm{E}}$, the time needed for 
a wavepacket of size $\lambda_{\mathrm{F}}$, the Fermi wavelength, to grow to the 
system size $L$ and delocalize.  Under chaotic dynamics, 
with Lypunov exponent $\lambda$, we have 
$\tau_{\mathrm{E}}\approx \lambda^{-1}\ln\left(L/\lambda_{\mathrm{F}}\right)$
and when no longer small compared to the dwell time, RMT 
stops being applicable.  However, such Ehrenfest time effects have been
incorporated into the semiclassical framework for all diagrams at leading order 
and some subleading order diagrams for low moments \cite{wj06,*jw06,br06,*br06b,wk10,wkr11}.  
Our systematic approach may then be useful beyond the universal regime.

\esec{RMT results}RMT provides the joint probability distribution of
the transmission eigenvalues \cite{beenakker97} which can be
integrated to obtain the transport moments, as was performed for the
conductance and its variance \cite{bm94,jpb94}.  Other quantities were
limited to diagrammatic expansions \cite{bb96} until the connection to
the Selberg integral was explored \cite{ss06,ssw08}.  Since then there
has been much interest and success in calculating the moments $M_n$ from the
circular ensembles \cite{vv08,novaes08,lv11,ms11,*ms11b}.

\esec{Semiclassical diagrams}Semiclassically, the elements $t_{oi}$ of the
scattering matrix are approximated \cite{miller75,richter00,rs02} by a sum over
the trajectories $\gamma$ which start in channel $i$ in one lead and end in
channel $o$ of the other.  They contribute a phase $\exp(\rmi S_{\gamma}/\hbar)$
with their action $S_{\gamma}$ so that $M_n$ is approximated by a sum over $2n$
trajectories of which half contribute with positive action and travel from
channels $i_j$ to $o_j$ while the other half contribute with negative action and
travel from channels $i_{j+1}$ to $o_j$ (we identify $i_{n+1}$ with $i_1$). 
Geometrically, if we reverse the direction of the trajectories with negative
action, the trajectories would form a single cycle visiting $i_1,o_1,i_2,\ldots$
in turn, as in \fref{semidiags}(a).  The phase involving the actions oscillates
in the semiclassical limit unless the total action difference is small on the
scale of $\hbar$.  To obtain the statistical properties of the moments we
average over a range of energies so that oscillating phases wash out and only
trajectory sets which achieve this small action difference contribute
consistently.  These, as for closed systems, come close in encounters 
while being nearly identical elsewhere (in `links'), as in \fref{semidiags}(b).

For $M_1$ (conductance), we have trajectory pairs starting and ending
together, and the contributing semiclassical diagrams were identified and
treated \cite{heusleretal06} precisely by cutting open the periodic orbit
diagrams of the form factor.  For $M_2$, which is related to the shot noise, the
contributing trajectory quadruplets can be formed by cutting periodic orbits
twice \cite{mulleretal07}, and this approach has been generalised in the
complementary work of Ref.\ \cite{novaes11}.  

\begin{figure*}
\includegraphics[width=\textwidth]{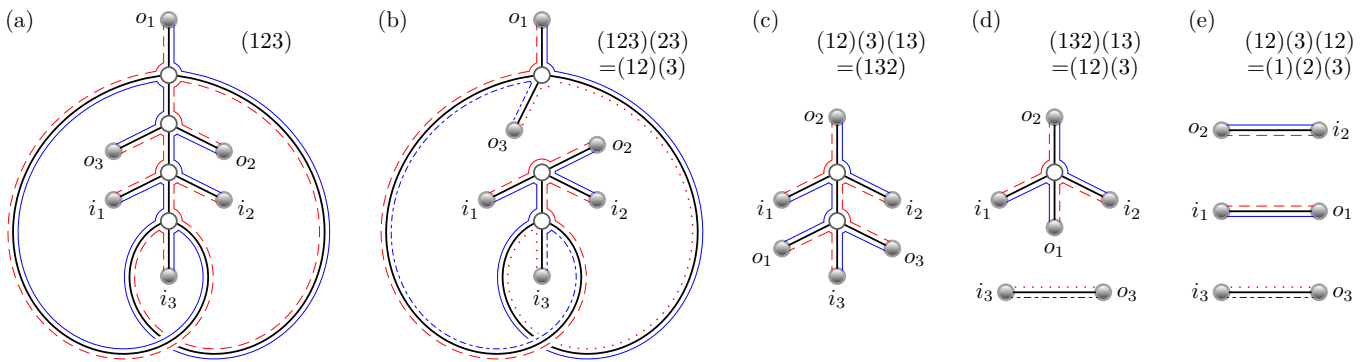}
\caption{\label{graphperm}(Color online) The trajectory sextet in (a) form a
single cycle or the permutation (123).  Untying the node with leaves $o_3$ and
$o_2$ (keeping $o_3$ connected to $i_3$) breaks the single cycle into two and
corresponds to multiplying (123) by (23) to obtain (12)(3) as in (b).  Repeating
the untying we move through (c) and (d) until we reach separated links in (e) or
the identity permutation.  Inverting the steps we can represent the original
diagram in (a) as the primitive factorization (12)(13)(13)(23)=(123).}
\end{figure*}

The contribution of each semiclassical diagram is a simple product of
its constituent parts \cite{mulleretal07}.  If the leads carry $\NL$
and $\NR$ channels respectively (with a total $N=\NL+\NR$) then each
incoming channel provides the factor $\NL$ and each outgoing channel
$\NR$.  More importantly, each link provides a factor of $1/N$ and
each encounter $-N$.  Assuming $\NL\sim\NR$, the order in $N^{-1}$ of
each diagram is the difference between the number of links and the
number of encounters (and channels).  The leading order diagrams can
then be redrawn as trees, or rather as paths around the tree along the
so-called `boundary walk' \cite{bhn08}.  The encounters are untwisted
to become roundabout nodes, the links edges and the channels leaves,
so \fref{semidiags}(b) morphs to \fref{semidiags}(c).  The leading
order of all $M_n$ was obtained in \cite{bhn08} by recursively generating
the trees.  Higher order diagrams involve closed cycles and a
graphical representation provided moment generating functions at the next two
orders \cite{bk11} which match an asymptotic expansion of RMT results
\cite{ms11}.  We now build on this to show exact concordance between
semiclassics and RMT for all moments and to all orders.  

\esec{Cancellations}First we show that the vast majority of possible
semiclassical diagrams cancel.  Since each encounter leads to a minus
factor, we will pair up diagrams that differ by one encounter and one
link to mutually cancel before counting the surviving
diagrams.  The pairing is realized using the following recursive
procedure:
\newcounter{steps}
\minlist
\begin{enumerate}
 \item Find the outgoing leaf $o_m$ (attached to an encounter node) 
   with maximal $m$.
\minlist
 \item If $o_m$ is attached to a node of degree 4 whose opposite edge ends in a
leaf, untie the node.
\setcounter{steps}{\value{enumi}}
\end{enumerate}
\minlist
To untie a node we break it into two parts keeping the path connecting
$i_m$ to $o_m$ intact.  This may separate a link directly connecting
$i_m$ to $o_m$ from the rest of the diagram.  Such a link is removed
from further consideration.  We repeat steps 1 and 2 while it remains
possible.  For example, from \fref{graphcancel}(a) we separate the top
encounter into two parts to give us \fref{graphcancel}(b) and
reduce $m$ to 1.

Once we can no longer perform step 2, we perform either of the following steps:
\minlist
\begin{enumerate}
\addtocounter{enumi}{\value{steps}}
 \item If $o_m$ is attached to a node of degree 4 whose opposite edge ends in
another encounter node, shrink the edge and join the two nodes together.
\minlist
 \item Otherwise, separate $o_m$ and its two neighbors from the encounter by
inserting a new link.
\setcounter{steps}{\value{enumi}}
\end{enumerate}
\minlist
These two operations are inverses of each other and provide
the required difference in the number of encounters [see
\frefs{graphcancel}(b) and (c)].  Finally we:
\minlist
\begin{enumerate}
\addtocounter{enumi}{\value{steps}}
 \item reverse all the operations performed at step 2.
\setcounter{steps}{\value{enumi}}
\end{enumerate}
\minlist
We thus reconstruct a diagram paired to the original one,
with a contribution of opposite sign.  All diagrams that ever arrive
at step 3 or 4 cancel with their partner, as, for example, the
diagrams in \frefs{graphcancel}(a) and (d).

\esec{Factorizations of permutations}Diagrams that never reach step 3
or 4 can only involve encounter nodes of degree 4 and, following steps
1 and 2 repeatedly, must eventually end up as a set of independent
links connecting each $i_j$ to its $o_j$.  For $M_n$ we initially have
trajectories along the boundary walk visiting the channels $i_1\to
o_1\to i_2 \ldots o_n \to i_1$ which we can represent as the cyclic
permutation $\sigma_n=(12\ldots n)$.  For systems without time
reversal symmetry, when we arrive at step 2 for the first time we must
have some $o_j$ opposite $o_n$.  The operation of untying is
equivalent to multiplying $\sigma_n$ by the transposition $(j\,n)$,
breaking the boundary walk into two cycles, $\sigma_n(j\,n)=(1\ldots
j)(j+1\ldots n)$.  Repeating steps 1 and 2 we multiply repeatedly on
the right by the pair of $o$ channel labels at each step 2 until we
arrive at the independent links whose boundary walk is the identity
permutation.  Reversing the untying of the nodes, we obtain a
factorization of $\sigma_n$ in terms of transpositions, which
represents the original diagram.  This process is illustrated in
\fref{graphperm}.  Because we always chose the $o_m$ with the maximal
$m$ at each step 1, the resulting factorization
$(s_1\,t_1)\ldots(s_{d}\,t_{d})$ can be written so that $t_j>s_j$ and
$t_k\geq t_j$ for all $k\geq j$.  Such a factorization is called
`primitive' and its depth $d$ is the number of nodes untied at step 2,
each of which removes two links and one encounter.  The number of
encounters in the diagram is equal to $d$ and the number of links
$n+2d$.  Diagrams without time reversal symmetry that survive 
the cancellation are therefore labeled by primitive factorizations 
and provide the semiclassical contribution $(-1)^{d}\NL^{n}\NR^{n}/N^{n+d}$.

\esec{Encounters in the lead}One complication is encounters that occur
in the leads.  For example, we could push the encounter in
\fref{semidiags}(b) to the left into the lead so that the incoming
channels coincide $i_1=i_2$.  We then lose two links, the encounter
itself and one channel so that the new diagram still contributes at
the same order but as $\NL\NR^{2}/N^3$ instead of
$-\NL^{2}\NR^{2}/N^4$. Likewise we could move the encounter to the
right until the outgoing channels coincide and the full contribution
of diagrams related to \fref{semidiags}(b) is the sum of these three
possibilities.

\begin{figure}
\includegraphics[width=\columnwidth]{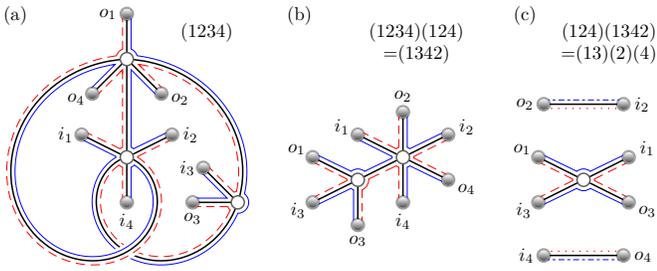}
\caption{\label{graphtouchleads}(Color online) The top node in (a) may move into
the outgoing lead which, by untying, we can represent as the permutation
(1234)(124)=(1342) and the boundary walk in (b).  Also placing the central node
in (a) in the incoming lead means we further multiply on the left by (124) to
obtain (c).}
\end{figure}

Whether encounters can be placed in the lead can be seen directly from
the graphical representation.  If every alternate link of a node ends
in an $i$ leaf, then the encounter can be placed in the incoming lead
and similarly for the outgoing lead.  Once encounters are in the
leads, they can no longer be joined or separated as in steps 3 and 4.
Instead, we consider the encounters in the leads as being already untied. 
Then we perform the procedure above to identify canceling
pairs.  Each node in the outgoing lead corresponds to multiplying
$\sigma_n$ on the right by the cycle of the labels of the $o$ leaves
(read off clockwise).  Nodes placed in the incoming lead multiply the
permutation on the left by the cycle of their $i$ labels, illustrated
in \fref{graphtouchleads}.  For the resulting permutation all
non-canceling diagrams are again labeled by the primitive
factorizations of the permutation in question.

\esec{Equivalence with RMT} The semiclassical result is
\begin{equation} \label{semimomentseqn}
M_n=\sum_{\chi_i,\chi_o}\frac{\NL^{c(\chi_i)}\NR^{c(\chi_o)}}{N^{n}}\sum_{d}
\frac{(-1)^{d}p_d(\chi_i\sigma_n\chi_o)}{N^{d}} 
\end{equation}
where $p_d(\cdot)$ is the number of primitive factorizations of depth
$d$, and $c(\cdot)$ the number of cycles.  In the sum over all
possible permutations $\chi_i$ and $\chi_o$, their cycles represent
incoming and outgoing channels which coincide.  For example, the
diagram in \fref{graphtouchleads}(a) contributes to the fourth moment
when the top and central nodes are placed in the leads,
$\chi_i=\chi_o=(124)(3)$.  The resulting primitive factorization (13)
leads to the contribution $-\NL^{2}\NR^{2}/N^{5}$.
For $M_2$, the only primitive factorizations are $(12)^{d}$ with odd 
and even $d$ being factorizations of $(12)$ and $(1)(2)$ respectively, 
which are also the two possibilities for $\chi_i$ and $\chi_o$.  This 
gives $M_2=\NL\NR/N -\NL^2\NR^2/(N^3-N)$.

The random matrix result, on the other hand, can be written exactly as
in \eref{semimomentseqn} but replacing
$\sum_{d}(-1)^{d}p_d(\pi)/N^{n+d}$ by coefficients $V(\pi)$; see
\cite{samuel80,mello90,bb96}.  That $V(\pi)$ is
the generating function for the number of primitive factorizations
$p_d(\pi)$ was recently established \cite{mn10} using an expression for
$V(\pi)$ in terms of characters of the symmetric group.
Here we sketch a simple alternative proof that is 
easy to generalize to systems with TRS.  The
$p_d(\pi)$ only depend on the cycle structure of $\pi$ so let
$c_1,\ldots,c_k$ be the lengths of the cycles in the permutation
$\pi=(12\ldots c_k)\cdots(n-c_1+1\ldots n)$.  First consider the case
when the term on the right of a primitive factorization of $\pi$ is of
the form $(s_d\,n)$.  Without this term, it is a factorization (of
depth $d-1$) of the permutation $\pi (s_d\,n)$.  If $s_d$ belongs to
the rightmost cycle of $\pi$ it splits into two, of lengths $q$ and
$r$ with $q+r=c_1$, while if $s_d$ belongs to cycle $j$ the rightmost
cycle joins with it to form a cycle of length $c_1+c_j$.  Finally, the
last term $(s_d\,t_d)$ can have $t_d \neq n$ only if $c_1=1$ and the
factorization does not have any occurrence of $n$ in it.  In this case
it is also a factorization of the permutation on $n-1$ elements with
cycle lengths $c_2, \ldots, c_k$.
In total we have
\begin{eqnarray}
&& p_d(c_1,\ldots,c_k) = \delta_{c_1,1}p_{d}(c_2,\ldots,c_k) \nonumber \\
&& +\sum_{c_j}c_jp_{d-1}(c_1+c_j,\ldots)
+\sum_{q,r=c_1}p_{d-1}(q,r,c_2,\ldots,c_k) \nonumber
\end{eqnarray}
exactly mirroring the recursion relations of $V$ \cite{samuel80}.

\esec{Time reversal symmetry}With TRS the
semiclassical diagrams are more complicated \cite{bk11}, and the
relevant combinatorial objects are permutations on $2n$ elements
$\bar{n}, \ldots, \bar{1}, 1, \ldots, n$.  The starting permutation
with no encounters in leads is encoded by
$\tilde{\sigma}_n=(\bar{n}\ldots\bar{1})(1\ldots n)$.  The
cancellation procedure remains the same, but now it is also possible
to untie nodes which have an $i_j$ leaf opposite $o_m$, which we
represent by multiplication on the right by $(\bar{j}\,m)$ and on the
left by $(\bar{m}\,j)$ [untying $o_j$ and $o_m$ also multiplies by
$(\bar{m}\,\bar{j})$ on the left in addition to $(j\,m)$ on the
right].  The non-canceling factorizations are then of the form
$\tilde{p}_d(\tilde{\sigma}_n)=(\bar{t}_d\,\bar{s}_d)\cdots(\bar{t}_1\,\bar{s}
_1)(s_1\,t_1)\cdots(s_d\,t_d)$ with $t_j>s_j$ and $t_k\geq t_j,
\bar{t}_j, \bar{s}_j$ for all $k\geq j$.

Since the two leads for the transmission are separate, still only
nodes with alternating $i$ or $o$ leaves can move into the leads and
we obtain a result like \eref{semimomentseqn} but involving the
doubled permutations.  Following similar reasoning
to above, the $\tilde{p}_d({\tilde{\pi}})$ satisfy the same recursion
relations as the coefficients $V$ from the circular orthogonal
ensemble so the moments are identical.

\esec{Spin-orbit interaction}The semiclassical framework includes spin-orbit 
interaction through an additional trace of a product of spin propagators 
along the trajectories \cite{zfr05,*zfr05b}.  The structure 
of the leading order diagrams makes this product identity and effectively leaves the leading 
order of $M_n$ unchanged \cite{wb07,*waltnerpc}.  For each order higher 
in inverse channel number, the chaotic spin-orbit interaction provides an additional 
factor of $-1/2$ (for spin $1/2$) compared to the contributions with TRS 
\cite{wb07,*waltnerpc}.  The cancellation procedure then still holds, 
and the effect can be included by simply substituting $\NL \to -2\NL$, $\NR\to-2\NR$ 
and multiplying by $-1/2$.  This is the same mapping as between the orthogonal and symplectic 
RMT ensembles \cite{bb96} so the moments are again identical. 

\esec{Conclusions}Though we focused on the moments of the transmission
eigenvalues, the combinatorial treatment here extends to non-linear
statistics by simply changing the starting permutation $\sigma_n$.
Moreover, since we have an exact concordance between the semiclassical
and RMT \cite{vv08,novaes08,lv11,ms11,*ms11b} moments of the
transmission eigenvalues, we obtain their probability distribution
semiclassically.  Indirectly, any RMT result derived from this
distribution, for example the nonlinear statistics in \cite{ssw08} and
the moments of the conductance and shot noise
\cite{novaes08,ok08,*ok09,kss09}, is now rooted in the chaotic
dynamics inside the cavity and the correlations between scattering
trajectories.

For energy dependent correlation functions, diagrams related to each other 
by the cancellation procedure no longer cancel exactly.  For the related quantities 
of Andreev billiards and Wigner delay times, the agreement between semiclassics and 
RMT remains limited to leading \cite{kuipersetal10,*kuipersetal11,melsenetal96,*melsenetal97} 
and several subleading orders \cite{bk10,bk11,ms11,*ms11b} respectively.

Finally, with the close connection between ballistic chaotic systems, RMT and
systems with weak disorder, the combinatorial ideas here should have parallels
in the diagrammatic perturbation theory of disorder.  They also form the start
point of including Ehrenfest time effects \cite{wj06,*jw06,br06,*br06b,wk10,wkr11} beyond the RMT
regime.

\vspace{\baselineskip}
\begin{acknowledgments}
\esec{Acknowledgments}We would like to thank Juan Diego Urbina and Klaus Richter for helpful comments 
and discussion as well as Marcel Novaes for sharing his alternative method and 
results \cite{novaes11}.  GB is funded by NSF award DMS-0907968 while JK acknowledges 
funding from the DFG through research unit FOR760.
\end{acknowledgments}

\end{document}